\documentclass[%
 aip,
 jmp,%
 amsmath,amssymb,
 reprint,%
]{revtex4-1}

\usepackage{graphicx}
\usepackage{dcolumn}
\usepackage{bm}

\begin{document}


\title[Enhanced longitudinal mode spacing in blue-violet InGaN semiconductor laser]{Enhanced longitudinal mode spacing in blue-violet InGaN semiconductor laser}

\author{I.V. Smetanin}
 \email{smetanin@sci.lebedev.ru}
 \affiliation{P. N. Lebedev Physical Institute,
Leninskii prospect 53, Moscow 119991, Russia}

\author{P. P. Vasil'ev}
\affiliation{P. N. Lebedev Physical Institute,
Leninskii prospect 53, Moscow 119991, Russia}
\affiliation{Centre for Photonic Systems, University of Cambridge, 9 J.J. Thomson Avenue, Cambridge CB3 0AF, UK}

\date{\today}

\begin{abstract}

A novel explanation of observed enhanced longitudinal mode spacing in InGaN semiconductor lasers has been proposed. It has been demonstrated that e-h plasma oscillations, which can exist in the laser active layer at certain driving conditions, are responsible for mode clustering effect. The resonant excitation of the plasma oscillations occurs due to longitudinal mode beating. The separation of mode clusters is typically by an order of magnitude larger that the individual mode spacing.

\end{abstract}

\maketitle

Blue-violet GaN-based diode lasers have attracted a great deal of attention over the last decade because of their growing applications in optoelectronics, optical storage systems, medicine, and some other field of modern science and technology. At the same time, many fundamental properties of this type of semiconductor lasers have not been completely investigated and are not yet well understood. One of the notable features of InGaN lasers is that mode spacing often experimentally observed in optical spectra of lasing emission can be 8-10 times larger as compared to that calculated from the cavity length \cite{Nakamura-1,Nakamura-2,Ropas,Eichler}. A few physical mechanisms have been proposed for explaining this effect. Irregular longitudinal mode spectra can be caused either by variation of the imaginary part of the refractive index or by a modulation of the real part of the refractive index along the active region. Irregularity may be attributed to interference caused by scattering  from randomly distributed scattering centers along the waveguide. In InGaN semiconductor lasers an inhomogeneous distribution of threading dislocations  may cause inhomogeneous strain and local scattering centers.

Meyer et al have  recently demonstrated \cite{Meyer} that optical gain fluctuations can be responsible for an additional filtering effect of longitudinal modes. Indeed, there are several mechanisms that can cause a decrease or increase or of the gain for  some longitudinal modes. Spatial fluctuations of the quantum well depth have been proposed  as one of the reasons. The gain of an individual longitudinal mode is then given by the overlap of the mode intensity distribution and the gain profile along the waveguide \cite{Meyer}. As a result, variations of modal gain in the laser cavity can amplify certain longitudinal modes more efficiently than another. This mechanism explained reasonably well optical spectra of InGaN lasers grown on SiC substrates, but failed to do the same for lasers grown on GaN substrates. On the other hand, at appropriate conditions temporal variations of the optical gain may also be responsible for irregular amplification of longitudinal modes.

\begin{figure}
\includegraphics{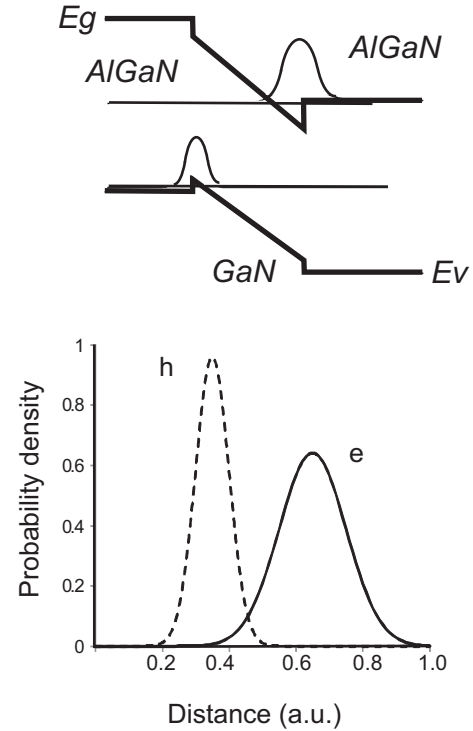}
\caption{\label{fig:epsart} Schematic of the band diagram  (top) and spatial distribution of the e-h probability densities in GaN quantum well. }
\end{figure}

In this letter, we have proposed a new physical mechanism explaining the mode spacing irregularity in GaN-based diode lasers, which is based on the resonant excitation of plasma oscillations in the coupled 2D layers of electrons and holes. A characteristic feature of GaN-based lasers (as compared to, say, GaAs/AlGaAs lasers) is the presence of an internal electric field caused by piezoelectric polarization existing in the structure \cite{Park}. As a result, the potential profile of the quantum well (QW) is distorted and the wavefunctions of electrons and holes localize at the opposite sides of the QW, see Fig. 1. On other words, an effective capacitor is formed inside the active layer, whose electrical field tends to compensate the strong intrinsic electric field. One has the two layers with the opposite charge of 2D electron and hole plasmas separated in space by a gap $\Delta$. The value of $\Delta$ depends on the internal field strength \cite{Fedorov}, being around the QW width when the field is strong and vanishing when internal field disappears (i.e., becomes compensated by the space-charge field). Separation of charged layers provides the low-frequency plasma eigenmode in such a system of coupled oscillators, which  vanishes for complete separation $\Delta \rightarrow \infty$  or merging of layers $\Delta \rightarrow 0$. The eigenfrequency for this mode in GaN is in sub-THz domain which provides the above experimentally observed modulation of lasing spectrum.

Plasma oscillations in the active layer of InGaN lasers can be excited by the slow traveling wave of photoinduced modulation of plasma density \cite{Smetanin}. Indeed, two longitudinal modes with intensities $P_n$ and $P_m$ , frequencies $\omega_n$  and $\omega_m$, and wave numbers $k_n$  and $k_m$ form the low-frequency beating wave with the wave number $k=k_n +k_m$  and frequency $\omega=\omega_n +\omega_m$. The rate of photoinduced electron - hole generation is then modulated as $S_{nm}=\sigma \sqrt{P_n P_m}\exp(i[kx-\omega t])+c.c.$, the coefficient $\sigma$ is proportional to the pair photo-production cross-section, and $x$ is the co-ordinate along the active layer.

Within the hydrodynamic approach, plasma oscillations in the system of two separated 2D layers of electrons and holes are described by the following system of equations \cite{Smetanin, Fetter}
\begin{eqnarray}
&&\frac{\partial n_e}{\partial t}+\frac{\partial n_e v_e}{\partial x}=-\gamma_R (n_e-n_{0e}) +S_{nm}, %
\nonumber \\
&&\frac{\partial n_h}{\partial t}+\frac{\partial n_h v_h}{\partial x}=-\gamma_R (n_h-n_{0h}) +S_{nm}, %
 \\
&&\frac{\partial v_e}{\partial t}+v_e\frac{\partial v_e}{\partial x}=-\frac{s_e^2}{n_{0e}}\frac{\partial n_e}{\partial x} -\frac{e}{m_e}(E_{11}+E_{21})-\gamma_e v_e, \nonumber \\
&&\frac{\partial v_e}{\partial t}+v_e\frac{\partial v_e}{\partial x}
=-\frac{s_h^2}{n_{0h}}\frac{\partial n_h}{\partial x} +\frac{e}{m_h}(E_{22}+E_{12})-\gamma_h v_h.\nonumber
\end{eqnarray}
Here, $n_{e,h}$, $v_{e,h}$ are the density and velocity of electron and hole 2D liquids, $n_{0e}$ and $n_{0h}$ are their steady-state densities, $\gamma_{e,h}$  are the characteristic rates of momentum relaxation of electrons and holes, $\gamma_R$ is the radiative lifetime of the carriers, $s_{e,h}$ is the speed of sound in 2D liquids. The first two equations are just the continuity equations of 2D e-h densities with the source term being $S_{nm}$. The second pair of the equations is the equations of motion of electrons and holes in self-consistent electric field. The 2D layer of electrons is affected by its intrinsic electric field $E_{11}$ and inter-layer electric field from the hole layer $E_{21}$ . Likewise, the layer of holes is affected by $E_{22}$ and the field from the electrons $E_{12}$. Fourier components of the intra-layer self-consistent fields are\cite{Fetter}
\begin{equation}
E_{11}(\omega, k)=i\frac{2\pi e}{\varepsilon}n_e(\omega, k), \;\;\;\;E_{22}(\omega, k)=-i\frac{2\pi e}{\varepsilon}n_h(\omega, k)
\end{equation}
here $n_e(\omega, k)$  and $n_h(\omega, k)$  are the density components of electrons and holes and $\varepsilon$ is the dielectric permeability of the medium. For the inter-layer components one has the following relations\cite{Fetter}: $E_{12}(\omega, k)= E_{11}(\omega, k)e^{-k\Delta}$ and $E_{21}(\omega, k)= E_{22}(\omega, k)e^{-k\Delta}$.

At the conditions of the problem at hand, one can neglect the internal pressure terms \cite{Fetter} in Eqs.(1) and introduce the 2D plasma frequencies of electrons $\Omega_{pe}^2 = (2\pi n_{0e}e^2/\varepsilon m_e)k$ and holes $\Omega_{ph}^2 = (2\pi n_{0h}e^2/\varepsilon m_h)k$. In the first order of perturbation theory, one can find the response functions $f_{e,h}(\omega,k)=n_{e,h}(\omega,k)/S_{nm}$ for electron and hole plasma layers,

\begin{eqnarray}
f_e(\omega, k)=i\frac{D_h(\omega+i\gamma_e)-(\omega+i\gamma_h)\Omega_{pe}^2e^{-k \Delta}}{D_e D_h
-\Omega_{pe}^2\Omega_{ph}^2e^{-2k\Delta}},  %
\nonumber \\
f_h(\omega, k)=i\frac{D_e(\omega+i\gamma_h)-(\omega+i\gamma_e)\Omega_{ph}^2e^{-k\Delta}}{D_e D_h
-\Omega_{pe}^2\Omega_{ph}^2e^{-2k\Delta}}.
\end{eqnarray}

where $D_e = (\omega+i\gamma_R)(\omega+i\gamma_e)-\Omega_{pe}^2$ and $D_h = (\omega+i\gamma_R)(\omega+i\gamma_h)-\Omega_{ph}^2$ determine the single-layer electron and hole plasma oscillation spectrum, respectively.
	
The resonant frequencies of plasma response are determined by the denominator in Eq. (3). In the relaxation-free case, it yields the following dispersion equation \cite{Chaplik1,Chaplik2}
\begin{equation}
(\omega^2-\Omega_{pe}^2)(\omega^2-\Omega_{ph}^2)-\Omega_{pe}^2\Omega_{ph}^2e^{-2k\Delta}=0
\end{equation}
In the limit of $k\Delta \gg1$ there is no bound between the layers, and 2D electron and hole liquids oscillate separately with their own plasma frequencies. When layers coincide, the eigenfrequency is $(\Omega_{pe}^2+\Omega_{ph}^2)^{1/2}$ and the second root of Eq.(4) vanishes. In the case of our interest, when the layer separation is small but not equal to 0, i.e. $k\Delta \ll1$, two solutions exist: 1) the high frequency branch $\omega_{high}^2 \approx (\Omega_{pe}^2+\Omega_{ph}^2)-\omega_{low}^2$ and 2) the low frequency branch
\begin{equation}
\omega_{low}^2 \approx 2k\Delta \frac{\Omega_{pe}^2\Omega_{ph}^2}{\Omega_{pe}^2+\Omega_{ph}^2}
\end{equation}
With $\Delta\rightarrow 0$, the low-frequency mode degenerates to the zero solution.

The plasma oscillations result in modulation of the optical gain and consequently to an additional filtering of the longitudinal modes. The mode filtering mechanism here is exactly the same as for the effect of gain fluctuations \cite{e}, when variations of modal gain in the laser cavity amplify certain longitudinal modes more efficiently than another. Now the question is whether the predicted modulation of the laser spectrum can describe observed enhanced mode spacing. According to Eq. (5), assuming the equal densities of electrons and holes $n_{0e}\approx n_{0h}=n_{2D}$ and taking into account that $kc=(\omega_n+\omega_m)\sqrt{\varepsilon}\approx2\sqrt{\varepsilon}\omega_n$, for the GaN diode parameters $m_e \approx 0.2m_0$, $m_h \approx 0.8m_0$, and the radiation wavelength $\lambda =425$~nm, we have the estimate for the low-band plasma frequency $\omega_{low}\approx 2\pi \times 0.84THz \times \sqrt{\Delta[{\rm nm}]n_{2D}[{\rm cm}^{-2}]/10^{12}}$ and for the modulation of laser diode spectrum $\Delta \lambda/\lambda \sim\omega_{low}/\omega\approx 1.19\times 10^{-3}\sqrt{\Delta[{\rm nm}]n_{2D}[{\rm cm}^{-2}]/10^{12}}$.

\begin{figure}
\includegraphics{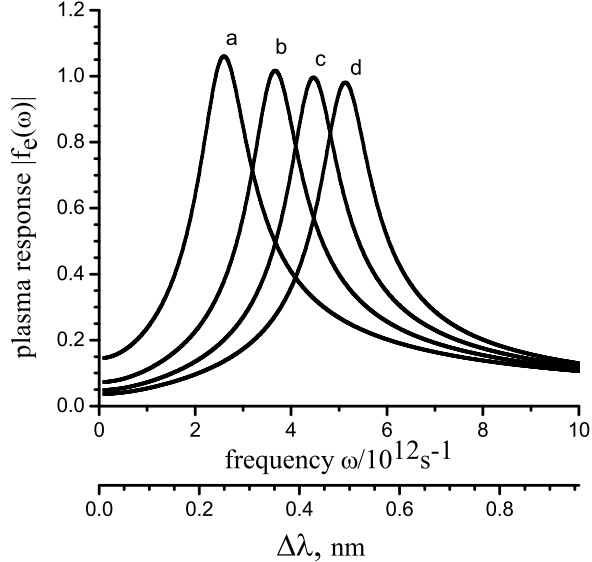}
\caption{\label{fig:epsart} Frequency response of the 2D plasma at a)$\Delta[{\rm nm}]n_{2D}[{\rm cm}^{-2}]/10^{12}=0.25$, b) 0.5, c) 0.75, and d)1.0}
\end{figure}

Figure 2 illustrates the dependence of the nonlinear response of the plasma double layer $|n_e(\omega)/S_{nm}|$ on the oscillation frequency $\omega$ which is described by Eq. (3). We use the above GaN diode parameters $\lambda=425$~nm , $m_e \approx 0.2m_0$, $m_h \approx 0.8m_0$, as well as the radiative relaxation rate $\gamma_R =10^9$~s$^{-1}$ , and the momentum relaxation rates $\gamma_e =\gamma_h=10^{12}$s$^{-1}$.

One can clearly see that the spectral modulation period $\Delta \lambda$ fits well to the observed experimental data \cite{Nakamura-1,Meyer}. Since the modulation period, which determines the observed mode spacing, depends on the carrier density and the charge layers separation in quantum well, different devices exhibit different $\Delta \lambda$. It is well-known that the carrier density in semiconductor lasers fixes at the threshold density and does not go up with an increase of the current above the lasing threshold. The threshold density depends on the cavity length, internal absorption, carrier lifetime and some other parameters. 

In conclusion, we present here a new explanation of the mode clustering effect which is generally observed in InGaN lasers. The internal piezoelectric field in these lasers results in a slight separation of wavefunctions of electrons and holes in the active region. We have demonstrated that in a bound two-layered e-h system plasma oscillations can exist. The plasma oscillations lead to the modulation of the optical gain in the cavity and consequently to an additional filtering of the longitudinal modes. Variations of modal gain in the laser cavity amplify certain longitudinal modes more efficiently than another. The estimation of the separation between the clusters of modes fits well to the previously observed experimental values.

This research is supported by the EC Seventh Framework Programme FP7/2007-2013 under
the Grant Agreement N 238556 (FEMTOBLUE).


\end{document}